\documentclass{aa}
\usepackage{graphicx}
\begin{document}
\title{Radio galaxies in cooling core clusters. Restarted activity in the nucleus of
3C\,317?}

\author{T. Venturi\inst{1}
          \and
          D. Dallacasa\inst{1,2}
          \and
          F. Stefanachi\inst{1}
          }
\offprints{tventuri@ira.cnr.it}
\institute{Istituto di Radioastronomia del CNR, Via Gobetti 101, 4129 Bologna, Italy
            \and
               Dipartimento di Astronomia, Via Ranzani 1, 40127 Bologna, Italy
              }
\date{Received \today; accepted ?}

\authorrunning{T. Venturi et al.}
\titlerunning{Radio galaxies in cooling core clusters: 3C317}

\abstract{We present the results of VLBA observations of the radio galaxy 3C\,317, 
associated with the cD galaxy UGC 09799 at the centre of the cooling core
cluster of galaxies A2052.\\
These observations were carried out at 1.7 GHz, 4.9 GHz and 8.3 GHz, in polarimetric 
mode, and allowed us to  image the parsec scale region of the source.
Our analysis suggests that the nucleus of 3C\,317 hosts a very young radio
source. The estimated radiative age
for the radio structure within the inner 10 pc is $\sim$ 170 yr. 
Given the existence of extended radio emission on the arcsecond scale, we suggest 
that 3C\,317 is a restarted radio galaxy. The implications of this result in the 
light of the interaction between radio plasma and thermal hot gas in clusters of
galaxies are briefly discussed.\\
\keywords{cooling flows - galaxies: clusters: individual(Abell2052) - galaxies:
 nuclei - galaxies: structure - radio continuum:galaxies
               }
}
\maketitle

\section{The current knowledge of 3C317 properties}
The radio galaxy 3C\,317 (RA$_{J2000}=15^h 16^m 44.5^s$, 
DEC$_{J2000}= 07^{\circ}01'17''$, redshift z = 0.035) is associated with 
the cD galaxy UGC9799, located at the centre of the cooling core cluster 
of galaxies A2052.\\
With a total power P$_{tot}$(1.4 GHz) = 5.4$\times10^{24}$ W Hz$^{-1}$, 
and a core power P$_{core}$(5 GHz) =  2.2$\times10^{24}$ W Hz$^{-1}$,
the radio galaxy falls in the range of low luminosity radio 
galaxies\footnote{We will use H$_0$= 50 km s$^{-1}$ Mpc$^{-1}$, q$_0$= 0 and 
$S\propto \nu ^{-\alpha}$ throughout the paper. For 3C\,317 1 mas $\sim$ 1 pc.
We define $h$=H$_0$/100}.
Unlike the typical Faranoff--Riley Type I radio sources 
(FRI, Fanaroff \& Riley 1974), characterized by core, twin jets and lobes, 
3C\,317 shows an amorphous 
halo surrounding a bright core, coincident with the optical centre 
of the galaxy. This core--halo structure is common among
the steep spectrum compact cores in radio sources associated with the 
central galaxies in cooling core clusters\ (\cite{Burns}).

While the connection between amorphous steep spectrum radio galaxies and 
cooling clusters is an established phenomenon, their causal link is
still unclear. Radio studies of central cluster galaxies 
(O'Dea \& Baum 1987; Burns 1990; Taylor et al. 1994) show that
cooling clusters may host both extended twin--jet radio 
galaxies and amorphous sources, however radio galaxies of this latter 
class are found exclusively in a cooling cluster environment.
Examples of cooling clusters with an extended central radio galaxy are
A2029, A4059 (Taylor et al. 1994), A2199 (Ge \& Owen 1993), Hydra A 
(Taylor \& Perley 1993); compact radio galaxies similar to 3C\,317 are 
3C\,84 in the Perseus cluster (B\"ohringer et al. 1993), 
PKS\,0745--191 (Baum \& O'Dea 1991), 2A 0335+096 
(Sarazin et al. 1995). Baum \& O'Dea (1991) and Sarazin et al. (1995)
discussed possible forms of interaction between the 
radio sources and the cooling flow, i.e. disruption of the 
radio jets by the high--pressure ambient gas, buoyant effects on the
amorphous radio emission, cooling gas as power supplier for the central 
radio galaxy, and finally gas heating in the inner cluster region from
the central radio galaxy.\\
More recently, thanks to the high resolution imaging capabilities of the 
X--ray satellite Chandra, the interaction between the radio plasma and 
the hot gas in cooling clusters has become even clearer, and it is now
accepted that the intracluster gas and the central radio galaxy mutually
influence each other. Examples include Hydra A \ (\cite{McNamara}; \cite{David}), 
and Perseus \ (\cite{Fabian}). In both cases, there is an 
anticoincidence between the radio and X-ray emission. In particular, the radio 
lobes on scales of tens (or hundreds) of kpc 
are located in ``holes'' of the X--ray emission.\\
Major interaction between the radio plasma and the intracluster gas
is now established also in the case of 3C\,317. Comparison between
arcsecond radio images and Chandra images suggests that  
the radio source has swept the gas away from the centre of the cluster,
compressing it into bright shells surrounding two ``holes'' in the X-ray emission
\ (\cite{Blanton1}; \ \cite{Rizza}). 

3C\,317 was studied in detail with the Very Large Array (VLA) at arcsecond 
resolution at 90, 20, 6 and 3.6 cm (Zhao et al. 1993).
It is characterized by a compact core and bipolar radio emission 
oriented in the North-South direction. This radio emission is embedded in an
amorphous halo, which shows considerable substructure at high resolution. 
The integrated spectrum over the range 0.01 -- 8 GHz shows a 
break around 0.5 GHz, and it is well fitted by two power laws with 
$\alpha=1.4$ for $\nu >0.5$ GHz and $\alpha=0.8$ for $\nu <0.5$ GHz. These 
spectral features and the radio morphology suggest that diffusion, 
synchrotron losses and electron reacceleration are important in this radio halo.\\
Another peculiar property is the high integrated rotation measure,  
$RM\sim -800$ rad m$^{-2}$ \ (\cite{Taylor}). High RMs are common 
in radio galaxies at the centre of cooling core clusters, and they are 
usually explained as due to a magnetized external screen
\ (\cite{Ge}).\\
At milliarcsecond resolution the source shows a bent morphology, and the radio 
emission on this scale has a flat spectrum up to 5 GHz \ (\cite{Venturi}).
This suggests that the subarcsecond core is active now, and raises the problem 
of its relation to the amorphous steep spectrum halo.

In order to study the polarimetric properties of 3C\,317 on the
parsec scale, and to address the issue of the connection between
the parsec and kiloparsec scale components of the radio emission, 
we performed a multifrequency polarimetric study with the
Very Long Baseline Array (VLBA).
Our results are presented in this paper, which is organised as follows.
In \S 2 we describe the radio observations 
and data reduction; the images are presented in \S 3 and are analysed in
\S 4; the results are discussed in \S 5; conclusions are given in \S 5.\\

\section{Observations}
3C317 was observed on March 5, 1999 using the full VLBA array, one VLA 
antenna and the Effelsberg antenna, simultaneously at 1.7 GHz (L band), 
4.9 GHz (C band) and 8.3 GHz (X band), in polarimetric mode.
The total allocated time (9 hours) was equally split among the three bands. 
The 4.9 and 8.3 GHz observations were carried out using two well
separated sub--bands of 8 MHz each, in order to optimise the u--v coverage 
and to derive the Rotation Measure with four points.
Details of the observations are given in Table ~\ref{table:tab1},
where for each band  we report the bandwidth, the observing 
frequencies ($\nu_{IF1}$ and $\nu_{IF2}$), the total time on source, 
the minimum and maximum 
baseline.\\
The data were correlated at the VLBA correlator in Socorro (New Mexico). 
Standard a--priori calibration, self--calibration, imaging and analysis were 
carried out using the NRAO Astronomical Image Processing System (AIPS). 
The accuracy of the absolute flux density scale is of the order of 
$\sim$ 4 -- 5 \% at all frequencies.\\
Polarisation calibration and imaging were performed at 4.9 GHz and 8.3 GHz
following the general method described in Cotton (1993), and implemented 
into AIPS by means of a number of procedures. No polarisation imaging was 
performed at L band.
Instrumental polarisation corrections were determined from 
measurements of 0016+731, known to be unpolarised, while
observations of 3C286 were used to derive absolute polarisation 
position angle $\chi$. We used the sum of the Q and U CLEAN 
components obtained from a subset of short baselines, 
and compared our results with Jiang et al. (1996). 
The accuracy of the absolute orientation of $\chi$ is estimated of the
order of $2^{\circ} - 3^{\circ}$ at both C and X bands. 
For our analysis, we imaged the IF1 and IF2, for the Q and U components, 
separately.\\

%
\begin{table*}[htbp]
\caption{Parameters of the observations}
\begin{center}
\begin{tabular}{|*{7}{c|}}
\hline
Band & Bandwidth  & $\nu_{IF1}$ & $\nu_{IF2}$ & t & U$_{min}$ & U$_{max}$\\
     & MHz &  GHz  & GHz  & hr   & $M\lambda$ & $M\lambda$ \\
\hline
L & 8+8 & 1.658 & 1.667 & 2.30 & 0.15 & 48\\  
C & 8+8 & 4.860 & 4.995 & 2.30 & 0.44 & 142 \\
X & 8+8 & 8.213 & 8.421 & 2.30 & 0.73 & 283 \\
\hline
\end{tabular}
\label{table:tab1}
\end{center}
\end{table*}

\section{Parsec scale morphology}
\subsection{Total intensity images}

\begin{figure}
   \centering
   \vspace{330pt}
    \includegraphics{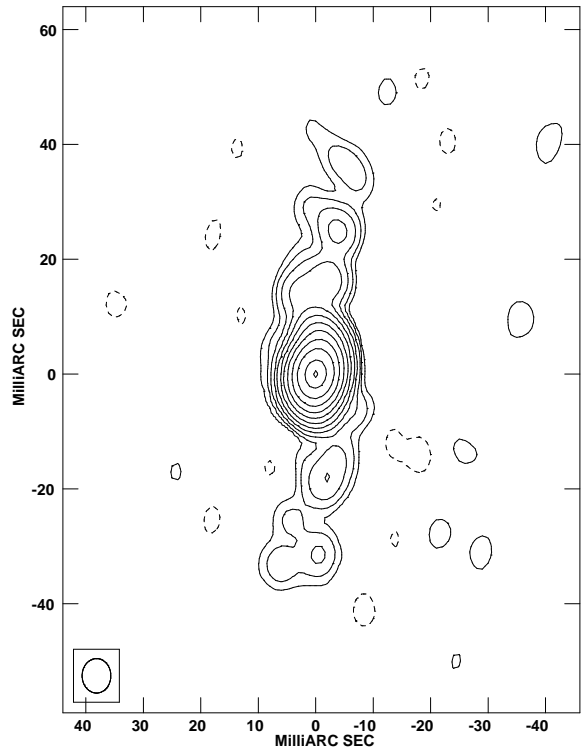}    
\caption{1.7 GHz image of 3C317. Peak: 276 mJy/beam. 
    Contours: 0.19 $\times $(--1, 1, 2, 4, 8, 16, 32, 64, 125, 250, 
500, 1000, 1400) mJy/beam. 
    Restoring beam FWHM: $ 6 \times 5 $ (mas) at $0^\circ$.}
       \label{fig:fig1}  
   \end{figure}

\begin{figure}
   \centering
   \vspace{330pt}
   \includegraphics{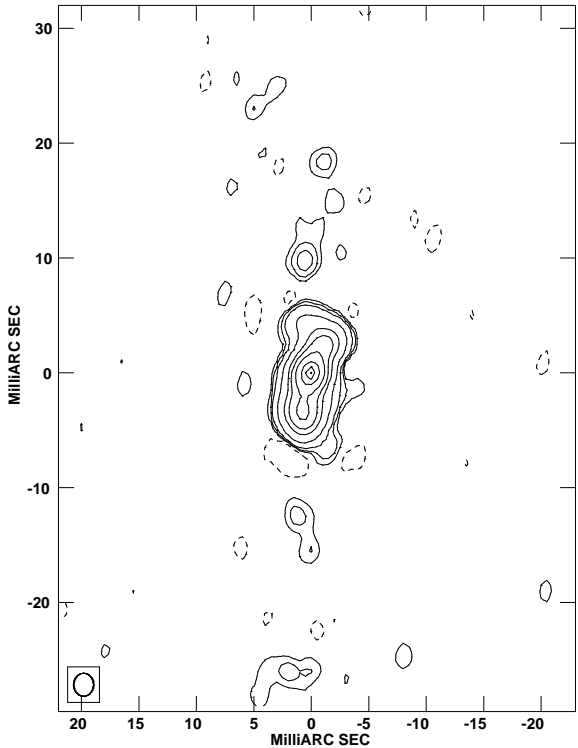}
   \caption{4.9 GHz image of 3C317. Peak: 155 mJy/beam. Contours: 
    0.19 $\times$ (--1, 1, 2, 4, 16, 64, 125, 250, 500, 700, 800) mJ/beam. 
    Restoring beam FWHM: $ 2 \times 1.73$ (mas) at $0^\circ$.}
       \label{fig:fig2}  
   \end{figure}

\begin{figure}
   \centering
   \vspace{330pt}
   \includegraphics{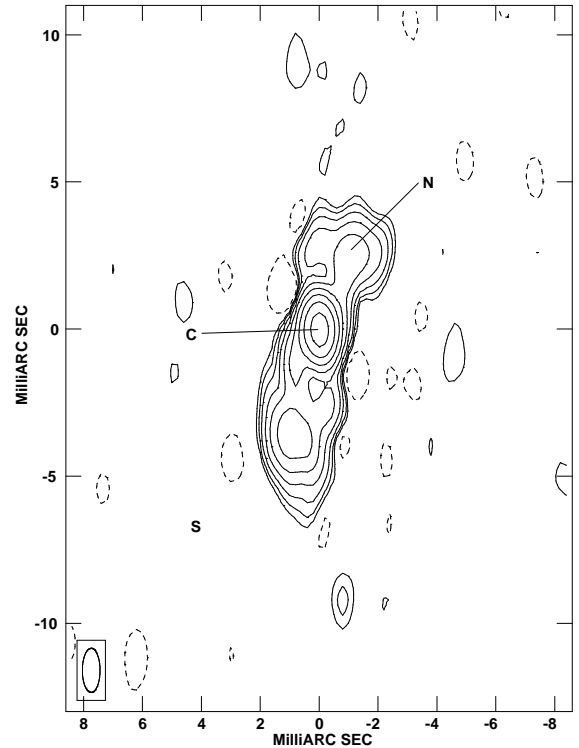}
   \caption{8.3 GHz image of 3C317. Peak: 102 mJy/beam. Contours: 
    0.27  $\times$ (--1, 1, 2, 4, 8, 16, 32, 64, 125, 250, 400) mJy/beam. 
    Restoring beam FWHM: $ 1.5 \times 0.6$ (mas) at $0^\circ$.}
       \label{fig:fig3}  
   \end{figure}

The uniform weighted total intensity images of 3C\,317 
at 1.7, 4.9 and 8.3 GHz are shown in Figs. \ref{fig:fig1}, \ref{fig:fig2} 
and \ref{fig:fig3} respectively, while
in Table ~\ref{table:tab2} we give the main parameters of the same images.
In particular for each band we report the total flux density, the peak flux 
density, the r.m.s. noise and the restoring beam.
The source is characterised by an overall symmetric structure, with a 
centre of activity and two opposing jets with twisted morphology. 
The brightness
asymmetry in the jets and the details of the twisting change with
frequency and resolution.

At 1.7 GHz (Fig. \ref{fig:fig1}) the source is dominated by a barely resolved 
central component, with two faint jets, elongated roughly in the N--S 
direction, whose surface brightness decreases quite smoothly. 
The two jets are slightly asymmetric, the northern being longer and 
brighter than the southern one.
The overall position angle is $\sim -5^{\circ}$. The morphology of 3C\,317
in this image resembles the bipolar structure of Zhao et al. (1993).
Hereinafter, we will refer to the jets on this scale as the bipolar 
emission.

The higher resolution of the 4.9 GHz image (Fig. \ref{fig:fig2}) allows us 
to resolve the 1.7 GHz central feature. The source has a symmetric 
structure, with a compact component, which includes the core
of the radio emission, and two short jets, bending at 
$ \sim 5$ mas from the peak, in a S--shaped structure. The overall position 
angle is $\sim -20^{\circ}$. At distance greater than 10 mas from the peak 
we find positive residuals on both sides, aligned in the North--South
direction, which we interpret as hints of the faint symmetric jets revealed at 
1.7 GHz, resolved out in this image.

At 8.3 GHz the S--shaped morphology of 3C\,317 is even more pronounced.
Two very short jets (see Fig. \ref{fig:fig3})
are aligned at $\sim -32^{\circ}$ with respect to the compact component
(C). They lose their collimation within the first few mas, to form two
features (N and S) which slightly bend counterclockwise, and give the 
source a ``twisted'' morphology.
These northern and sourthern features are very similar in shape, 
size and flux density. In particular, the ratios between their flux 
density, major and minor axes are respectively 1.5, 1.4 and 1.3. 
We note that the southern structure S is longer and brighter than the 
northern one. This is opposite to what is found at 1.7 GHz on the scale of 
tens of mas, where the northern jet is more prominent than the southern. 
The overall source structure at this frequency is still elongated roughly 
in the N--S direction, with a position angle of $-22.5^{\circ}$. 
At this frequency and resolution the faint jets revealed at 
1.7 GHz are completely lost.

We are aware of the fact that the images presented here differ 
from those (EVN and MERLIN) published in Venturi et al. (2000). 
The details of both sets of images can be accounted for by the 
different u--v coverage, resolution and sensitivity. 
Moreover, the combination of the EVN and MERLIN observations,
may have resulted in residual amplitude errors,
affecting the reliability of the images. 
Despite the much better quality of the present observations and 
the different spatial regions sampled in the source, however,
we note that the orientation of the inner parsec--scale 
structure is in reasonable agreement with the older datasets.

\medskip
As a general comment, a remarkable feature of the overall structure of 
3C\,317  is the change of the position 
angle, which smoothly rotates by $\sim 20^{\circ} - 25^{\circ}$ going 
from the bipolar low brightness jets detected at 1.7 GHz on a scale of 
20 -- 40 mas (corresponding to a linear scale of 20 to 40 pc) to the 
inner symmetric structure detected at 
8.3 GHz within 5 mas from the central compact component ($\sim$ 5 pc).
It is noteworthy that the length and brightness asymmetry of the
morphology changes from 1.7 GHz  to the inner mas imaged at 8.3 GHz.  
All these details reflect the complexity of the source on all scales. 
Even though
the overall orientation of the structure embedded in the radio halo is the 
North -- South direction, details and fine structure change considerably
depending on the frequency and resolution (Zhao et al. 1993; Ge \& Owen 1994;
Venturi et al. 2000).

The flux density ratio between the northern and sourthern jet components, 
as derived from the 8.3 GHz image, i.e. R = S$_S$/S$_N$ = 1.5,
gives us some information on the source orientation in the plane of the sky. 
Assuming that any brightness asymmetry is due to Doppler boosting in an 
intrinsically symmetric source, R = 1.5 implies $\beta$cos$\theta \sim$ 0.08,
where $\beta$ is the intrinsic speed of the relativistic plasma and
$\theta$ is the viewing angle. Such constraint gives very large angles
to the line of sight even for a mildly relativistic bulk motion of the
radio plasma, i.e. $\theta(\beta=1) \sim 85^{\circ}$, and 
$\theta(\beta=0.5) \sim 81^{\circ}$.


%


%


%

\subsection{Polarisation properties}

The arcsecond scale core of 3C317 is known to be unpolarised (Ge \& Owen
1994). The possibility of beam depolarisation is ruled out by our
observations: we did not manage to reveal any polarised emission from
3C317 either at C or ad X band. In particular at X band, where we had the
best combination of sensitivity and resolution, we can exclude any
polarised emission stonger than 0.3 mJy/beam at the full resolution of 
the observations.
In the polarised emission
images at 8.2 and 8.4 GHz, after correction for the ricean bias, the
r.m.s. noise is $\sim$0.06 mJy/beam. The very few 5$\sigma$ peaks 
(mostly off--source) happen at different locations at the two frequencies.
Therefore, either the source is completely unpolarised on the pc scale or the
Faraday screen is not resolved by the present observations.
We can set conservative upper limits to the fractional peak polarisation
from the jets  to 10\% for the northern one and to 5\% for the brighter
southern one. The core is at most 0.3\% polarised.
We obtained even more stringent upper limits (4\%, 2\% and 0.2\%
respectively) from the C band data, but we must consider that the lower
resolution does not allow to disentangle possible structure in the
polarised signal among the three main components partly blended together.


\begin{table*}[htbp]
\caption{Parameters of the images}
\begin{center}
\begin{tabular}{|*{5}{c|}}
\hline
Band & S$_{tot}$ & S$_{peak}$  & r.m.s. & Restoring FWHM and PA \\
     & mJy & mJy/beam  & $\mu$Jy/beam & mas, $^{\circ}$  \\
\hline
L & $353\pm14 $ & $276\pm11 $ & $64$  & 6.0 $\times$ 5.0 (mas) at 0 $^\circ $ \\  
C & $345\pm14 $ & $155\pm~6  $ & $65$ & 2.0 $\times$ 1.7 (mas) at 0 $^\circ $ \\
X & $217\pm9  $ & $102\pm~4  $ & $90$ & 1.5 $\times$ 0.6 (mas) at 0 $^\circ $ \\
\hline
\end{tabular}
\label{table:tab2}
\end{center}
\end{table*}


\section{Parsec scale properties}

\subsection{The parsec scale spectrum}

We used our multifrequency data to determine the spectral index on the 
parsec scale structure between 1.7 and 4.9 GHz, and between 4.9 and 8.3 GHz. 
For our analysis we made images of the source using the 
same maximum and minimum baseline in the u--v coverage, the same 
gridding and the same restoring beam.

Between 1.7 and 4.9 GHz the u--v cut applied to our datasets
allowed us to determine only the total spectral index of the brighest 
component in the 1.7 GHz image (Fig. \ref{fig:fig1}),
given that the resolved emission seen at 1.7 GHz is barely detected
at 4.9 GHz and the structure seen in this latter image is almost unresolved in 
the former. Our images were fitted with a gaussian component, whose dimensions 
are $4 \times 1 $ mas. The spectral index is $\alpha_{1.7}^{4.9} 
\sim -0.01_{-0.09}^{+0.10}$, in perfect agreement  with the values derived in 
the previous work of Venturi et al. (2000)  and Zhao et al. (1993).\\

In order to derive an estimate of the spectral index in the 
bipolar milliarcsecond jets visible at 1.7 GHz, we made a naturally
weighted 4.9 GHz image, and convolved it with the full resolution
1.7 GHz restoring beam. We are aware of the different coverage 
at the short baselines introduced by this procedure, however our
approach is justified by the fact that the full resolution 4.9
GHz image (Fig. \ref{fig:fig2}) clearly shows that the 1.7 GHz
bipolar emission exists but it is resolved out in this higher 
resolution image. We managed to image the extended structure of 
3C\,317 at 4.9 GHz, and obtained 
$\alpha_{1.7}^{4.9} \sim 0.77^{+0.09}_{-0.10}$.

The images made at 4.9 GHz and 8.3 GHz for the determination of the spectral
index are detailed enough to allow an estimate of $\alpha_{4.9}^{8.3}$ 
in the compact component C and a global value in the northern and southern region.
We found that $\alpha_{4.9}^{8.3}$(C)$\sim 0.7\pm 0.2$, while we found average 
values of 1.2$\pm 0.2$ and 1.1$\pm 0.2$ in the northern and southern region 
respectively.
Our analysis confirms that component C may host the ``core'' of
the radio emission.
The total spectral index was computed on the basis of the total
flux density in the two images, derived be means of the task TVSTAT.
We found $\alpha_{4.9}^{8.3} \sim 0.9 \pm 0.2$.

\begin{figure}
   \centering
   \vspace{300pt}
   \includegraphics{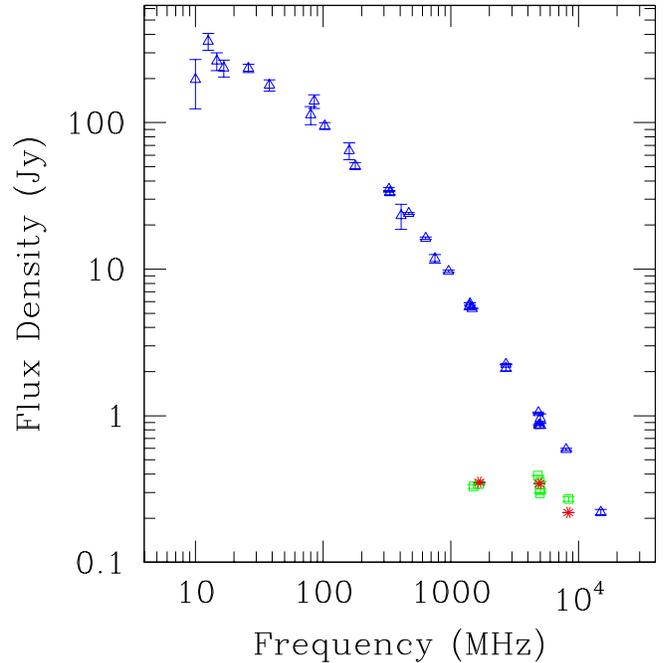}
   \caption{Spectrum of 3C317 from 10 MHz to 15 GHz. Red stars 
represent the VLBA data presented in this paper; open blue triangles 
refer to the total flux density measurements from the literature, 
open green squares are core flux density derived in previous works 
(see Venturi et al. 2000 for references).}
       \label{fig:fig4}  
   \end{figure}

In order to get a global picture of the radio spectrum of 3C\,317,
from the literature we collected the total flux density measurements, 
the arcsecond core data (Morganti et al. 1993, Zhao et al. 1993),
the milliarcsecond data (Venturi et al. 2000), and plotted them in
Fig. \ref{fig:fig4} together with the total flux density data from the
present work (see Table 2).\\
Our data are in excellent agreement with the previous observations on the
parsec and on the sub--arcsecond scale. Moreover, the addition of the parsec
scale 8.3 GHz flux density clearly shows the presence of a nuclear turnover 
$\nu_{to}$ between 1.7 and 5 GHz, ruling out the possibility of a flat
spectrum from 1.7 up to 8.3 GHz. We note that the nuclear spectrum 
shown in Fig. \ref{fig:fig4} is considerably steep
for  $\nu > \nu_{to}$. 
The values of the spectral index obtained for the various features in
the source are summarised in Table \ref{table:tab3}.\\
We are aware of the fact that the spectral index in the bipolar emission 
on the scale of 20 -- 40 pc is consideraby flatter than in the innermost 
regions. It is not clear how these values can be reconciled. One possibility is 
that the jet flow experiences some form of reacceleration on the
scale of $\sim$ 10 pc. Under this hypothesis, the lack of emission
beyond this distance in the 4.9 GHz image would then be due to an
increase in the jet transport efficiency, rather than to a surface brightness 
decrease as consequence of jet broadening. Alternatively, small oscillations 
of the symmetric jet structure in the direction of the line of sight may 
influence our total flux density measurements, thus affecting the
estimate of the spectral index. Unfortunately, our data and images are
not adequate to investigate any of these suggestions.


%
%
%
\begin{table}[htbp]
\caption{Spectral index of the components of the milliarcsecond structure of 
3C\,317}
\begin{center}
\begin{tabular}{|*{3}{c|}}
\hline

         &   $\alpha_{1.7}^{4.9}$  & $\alpha_{4.9}^{8.3}$ \\
\hline
Central component   &  --0.01  & +0.9   \\
Bipolar emission    &   +0.77  & --    \\
\hline
Component C         &    --    & +0.7   \\
Region N            &    --    & +1.2   \\
Region S            &    --    & +1.1   \\
\hline
\end{tabular}
\label{table:tab3}
\end{center}
\end{table}

\subsection{Analysis of the parsec scale spectrum}

\begin{figure}
   \centering
   \vspace{300pt}
   \includegraphics{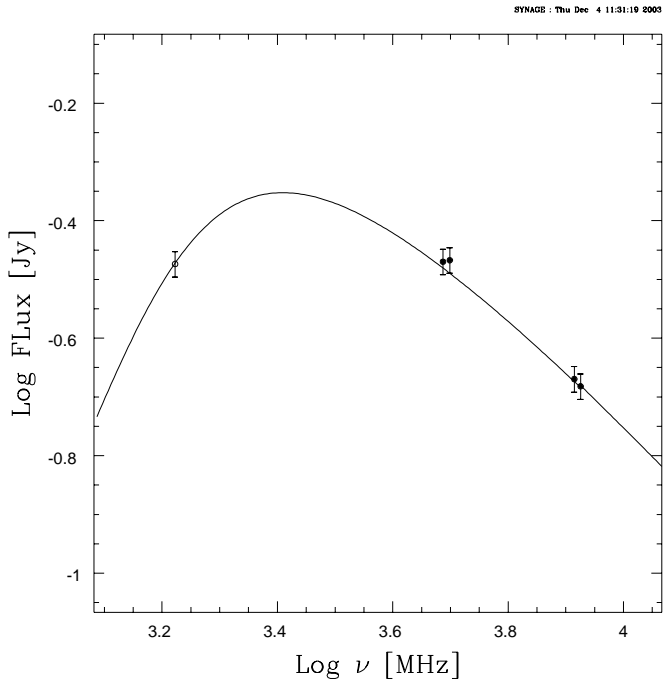}
   \caption{Best fit of the nuclear spectrum of 3C\,317 obtained with the program 
Synage using a ``continuous injection'' model.}
   \label{fig:fig5}  
   \end{figure}

We carried out a detailed study of the total parsec scale spectrum of
3C\,317, in order to derive an estimate of the break frequency $\nu_{br}$
and of the injection spectrum of the radiating electrons.
We used the most updated version of the program Synage 
(Murgia \& Fanti 1996), and found that the best fit to our
spectrum is given by a ``continuous injection'' model (Kardashev 1962),
with an initial electron injection index $\alpha_{inj} = 0.55$,
a break frequency $\nu_{br} \sim 6.5$ GHz and a turnover frequency 
$\nu_{to} \sim 2.6$ GHz (see Fig. \ref{fig:fig5}). We note that the
flux density values used for the best fit refer to the inner 10 mas
of 3C\,317, i.e. the bipolar emission at 1.7 GHz is not included.

\subsection{The physical conditions at equipartition}

In order to understand the nature of 3C\,317 and compare its parsec--scale 
properties to those of other classes of compact radio sources, for each
component (see Fig. 3) we computed the intrinsic parameters in the source, 
i.e. magnetic field, internal pressure and total energy, assuming that 
equipartition holds in the source.

We used the standard formulae in Pacholczyk (1970), assuming that the
relativistic particles and the magnetic field fully occupy the same volume 
($ \Phi = 1$), and that the amount of energy in heavy particles is the same as 
that in electrons ($k=1$). We integrated over the frequency range
$10^{7}- 10^{11}$ Hz, used an ellipsoidal geometry for each component
and the values for $\nu_{br}$ and $\alpha_{inj}$ derived from the study
of the nuclear spectrum.

For the nuclear region imaged at 8.3 GHz (component C, and the two short 
twisted jets) we obtained 
a global equipartition magnetic field $H_{eq} \sim 4.3 \times 10^{-2}$ Gauss, 
while in the extended bipolar emission
detected at 1.7 GHz we found an average value $H_{eq} \sim 2.4 \times 10^{-3}$ G.
The total energy is E$_{tot}\sim 9.5\times10^{51}$ and $\sim 3\times10^{52}$ erg
for the central region and for the bipolar emission respectively. The
corresponding minimum internal pressures are $P_{eq} \sim 1.7\times10^{-5}$ and 
$\sim 2\times10^{-7}$ dyn/cm$^2$.

\subsection{Considerations from the spectral analysis}

The nuclear properties of 3C\,317, as derived from our spectral
analysis, are similar to those found in Compact Steep Spectrum 
sources (CSS). \\
The spectrum of the nuclear region in 3C\,317, with a turnover at
$\nu_{to} \sim$ 2.6 GHz, and the steep spectrum for $\nu > \nu_{to}$ 
are reminiscent of the spectra
of Giga--Hertz Peaked--Spectrum (GPS) radio sources, as it can be clearly 
seen in O'Dea (1998) and in Stanghellini et al. (1998). Moreover,
other properties of the source show similarities with CSS and GPS 
sources. Beyond the double sided emission on the nuclear scale, the
lack of polarisation is consistent with the finding in Cotton et al. (2003) and
Fanti et al. (2004) for a sample of CSS radio sources: intrinsically small
radio sources appear to be unpolarised on scales smaller than a few kpc,
and the projected linear size at which the sources start to be
statistically polarised gets smaller with increasing frequency (Fanti et
al. 2004). In particular, the projected linear size (10 pc) of the
structure in Fig. 3 requires no polarisation at X band, as observed.
In summary, as it is the case for most of the CSS sources unpolarised on
the VLA scales, beam depolarisation is not effective.\\
Last but not least, our estimates for the equipartition parameters are in good 
agreement with the values found for other compact sources with comparable size
and radio power. The equipartition field values H$_{eq}$ found from VLBA 
observations  of the compact steep spectrum sources in the B3--VLA sample go up 
to 20 mG (Dallacasa et al. 2002), 
and their average value is of the order of 2 mG, as found in the 1.7 GHz bipolar 
emission of 3C\,317.

\section{Discussion}

The results of our multifrequency parsec--scale study of 3C\,317 
can be briefly summarised as follows.\\
{\it (a)} The source has a two--sided morphology, whose overall
position angle rotates by $\sim 20^{\circ} - 25^{\circ}$ going from
a few mas (8.3 GHz) out to a few tens of mas (1.7 GHz).
At 8.3 GHz 3C\,317 is characterised by a central component and two 
short symmetric jets, ending into two high brightness 
regions, with a twisted S--shaped morphology. In the following discussion 
we will refer to these two regions as to the {\it lobes}.
Constraints from the flux density 
asymmetry suggest that 3C\,317 lies almost in the plane of the sky, i.e.
it is viewed under an angle $\theta \sim 81^{\circ} - 85^{\circ}$;\\
{\it (b)} the total spectrum on the parsec scale is convex, with a
turnover $\nu_{to} \sim 2.6$ GHz, and a steep spectral index 
at frequencies $\nu > \nu_{to}$, i.e. $\alpha_{4.9}^{8.3} \sim 0.9$;\\
{\it (c)} the source is unpolarised on this scale, and the magnetic
field is randomly oriented;\\
{\it (d)} the average magnetic field at equipartition in the nuclear
region and in the extended bipolar emission (1.7 GHz image at the resolution 
6$\times$5 mas) is respectively H$_{eq} \sim$ 43 mG and 2.4 mG.
These values, and the other equipartition parameters, 
are in agreement with those found for GPS sources and for the most compact 
CSS sources.

The S--shaped morphology of 3C\,317 at high resolution is reminiscent of 
1946+708 (Taylor \& Vermeulen 1997) and 2352+495 (Wilkinson et al. 1994), 
classified as Compact Symetric Objects (CSOs, see O'Dea 1998 for a recent review).
This feature, together with points {\it (b), (c)} and {\it (d)} above,
suggest that the nuclear region of 3C\,317 itself may be a member of this class.

%
%

\subsection{Restarted activity in the nucleus of 3C\,317?}

It is now accepted that CSO radio sources are on average young 
objects, which will evolve into sources of large angular 
size, either FRI or FRII radio galaxies, depending on the 
initial radio power (Fanti et al. 1995; Readhead et al. 1996; 
Owsianik \& Conway 1998). In particular, the detection of hotspot 
proper motion in $\sim$ 12 CSOs (Polatidis et al. 2002; 
Polatidis \& Conway 2003), coupled with recent ``spectral ageing'' 
studies (Murgia 2003), strongly support the proposed evolutionary
scenario.

In order to test the suggestion that the nucleus of 3C\,317 is a CSO, 
we computed the equipartition field H$_{eq}$ in the lobes N and S of the 
high resolution 8.3 GHz image and used the formula by Murgia (2003)
to estimate the radiative age of the electrons in these regions
(see Fig. 3):

$$
t_{syn}(yr)=5.03\times 10^4 \times (H_{eq})^{-3/2} [(1+z)\nu_{br}]^{-1/2}
$$

\noindent
where $H_{eq}$ is in units of mG and $\nu_{br}$ in GHz. 
We could not carry out a detailed spectral study for the lobes, 
given the very small number of resolution elements along the structure,
therefore we assumed an upper limit for $\nu_{br}$ of the order
of 6.5 GHz, i.e. the break frequency for the whole nuclear region
(see Section 4.3). We estimated an equipartition field of 23.5 mG,  
and derived an average radiative age for the two lobes of the order 
$t_{syn}$ = 170 yr. This result suggests that the radio emission 
coming from the nuclear region very is young.\\

We used the radiative age to estimate the expected 
proper motion in the inner region of 3C\,317. Murgia (2003) showed that
the radiative ($t_{syn}$) and dynamic age (t$_{dyn}$) are in very good 
agreement in the only well studied case, i.e. B\,1943+546, so our approach 
is based on this result.
Assuming $t_{syn} = t_{dyn}$ = 170 yr, and a distance of $\sim 7$ pc
between the two lobes (see Fig. \ref{fig:fig3}), we obtained a
separation velocity $v_{sep} \sim 0.14~hc$. This value is in the range
of proper motions found in the hot spots of CSOs reported in 
Polatidis \& Conway (2003).

\medskip
The very young dynamical and radiative ages of CSOs poses the problem
of recurrent activity. 
Beyond the theoretical work of Reynolds \& Begelman (1997), 
hints on the fact that the life of a radio source may be characterized by 
alternate active and quiescent phases come from a number of observational
results. In particular, 
{\it (a)} the large angular size double--double radio galaxies 
(DDRGs) are interpreted as due to intermittent radio activity in the host 
galaxy (Schoenmakers et al. 2000; Lara et al. 1999); 
{\it (b)} the existence of a superluminal parsec--scale nucleus 
associated with an old extended radio galaxy has been found in B2\,1144+35 
(Giovannini et al. 1999) and in 3C\,338 (Giovannini et al. 1998), pointing
again in the direction of restarted radio activity; 
{\it (c)} finally, recurrent activity in central dominant cluster 
galaxies is suggested by the correlation between X--ray cavities and 
low brightness radio emission in galaxy clusters (McNamara et al. 2000).\\
In the case of CSOs, assuming that even a fraction of them represents a
renewed cycle of radio activity, old extended radio emission should
be found around some members of this class. This is indeed the case for
a few objects, such as 0108+388 (Owsianik et al. 1998), 0402+379 
(Maness et al. 2003) and 1245+676 (Marecki et al. 2003).

The nuclear properties of 3C\,317, coupled with 
the information on the arcsecond scale radio emission, and the 
high resolution X--ray Chandra images of the intracluster gas in
A\,2052, are consistent with the idea that the radio
activity in 3C\,317 is recurrent, and that at present the source is  
undergoing a newly born active cycle.\\
We estimated the age of the bipolar radio emission over $\sim$ 40 pc,
well visible at 1.7 GHz and detected also at 4.9 GHz. We note that 
we have no information on the break frequency $\nu_{br}$ for this
feature. Given our knowlegde on the inner CSO structure, we assumed
that $\nu_{br}~<~5$ GHz, and derived that the radiative age must
be $t_{syn} > 6\times10^3$ yr. Even though we should consider this result
only as indicative, due to the underlying assumptions on the break frequency,
we note that this lower limit is in reasonable agreement with the hypothesis 
made by many authors that the duration of the cycles of activity is in the 
range  $10^4 - 10^5$ yr (i.e. Reynolds \& Begelman 1997).\\
In order to complete our study of the radio emission age in 3C\,317, we 
fitted the spectrum of the extended arcsecond emission,
in order to derive the overall source age. 
The best fit to the spectrum is given by a continuous injection
model with self--synchrotron absorption, and
provides a break frequency $\nu_{br}$ = 780 MHz. We wish to point out that
the continuous injection model does not contraddict the proposed frame
of recurrent activity, since during the lifetime of the extended radio
emission the nucleus appears ``on average'' active.
If we assume an average magnetic field of $\sim 20 \mu$G (Zhao et al. 1993)
we obtain a global age t$_{syn} \sim 2\times 10^7$ yr.

\subsection{Alternative interpretations}

\subsubsection{Slow jets?}
An alternative hypotesis to recurrent bursts of radio activity in 3C\,317 is
the possibility that the intrinsic plasma speed in the jets is not relativistic,
and that major interaction with the external medium prevents them from a 
full development. Our high frequency and high resolution images clearly show
that no trace of collimated jets is visible beyond $\sim$ 10 mas from the peak.
The symmetry of the parsec--scale emission (as discussed in Section 3.1)
suggests that relativistic beaming is playing a small role for this source, 
therefore mild intrinsic speeds are not contraddicted by our results.\\
However, we know from the arcsecond emission that the plasma indeed reaches 
those scales, therefore relatively fresh electrons must be transported 
out on scales of tens of kpc, given that buoyancy is not effective to produce 
such extended emission on timescales of the order of 10$^7$ years.

\subsubsection{Particle reacceleration in the cooling flow}

Another intriguing possibility is that the extended radio emission
in 3C\,317 is a {\it mini--halo}.
Such radio sources are found around powerful radio galaxies
located at the centre of cooling core clusters.
Observationally, their extension is of the order of a few hundred kpc, 
and they are characterised by low brightness extended emission with a steep
spectrum ($\alpha~>$ 1). The prototypical example of this class of radio 
sources is 3C\,84 in Perseus (Burns et al. 1992).
Mini--halos have been recently explained in terms of reacceleration of 
old relativistic electrons, due to MHD turbulence in the cooling flow
(Gitti et al. 2002). \\
A comparison between the radio and X--ray emission in the Perseus cluster
and in A2052 is providing very interesting pieces of information.
The similarity of the radio/X--ray overlay in the inner region of the two 
clusters, i.e. within $\sim 1.5$ arcmin (Fabian et al. 2000 and 
Blanton et al. 2001, respectively for Perseus and A2052) is impressive. 
Furthermore, the 327 MHz overall morphology of the inner 10 arcmin in 
3C\,84 (Sijbring 1993) is identical to that of 3C\,317 at the same 
frequency (Zhao et al. 1993). From this image, the largest linear size 
of 3C\,317 is only $\sim$ 130 kpc, to be compared respectively to the 
$\sim$ 400 kpc and $\sim$ 200 kpc of the mini--halos in Persues and in 
A\,2626 (Gitti et al. 2004). We note however that the high resolution of 
the 327 MHz image in Zhao et al. (5$^{\prime\prime}\times5^{\prime\prime}$) 
does not allow a thorough comparison. Imaging at low frequency and 
resolution, and high sensitivity is necessary to properly study the case
of 3C\,317.\\
We are at present investigating the possibility that the extended emission
of 3C\,317 is also the result of reacceleration mechanisms due to the
central cooling flow.

\section{Conclusions}

Our polarimetric VLBA multifrequency observations of 3C\,317,
the central radio galaxy in the cooling core cluster A\,2052,  
revealed a compact double morphology over a scale of few mas.
The observational properties are consistent with the idea that
the nucleus of 3C\,317 is a young CSO. In particular, the peak in 
the radio spectrum at $\sim$ 2.6 GHz, the lack of polarisation, the 
orientation almost in the plane of the sky, and the values of the intrinsic
physical parameters are common to CSS and to GPS sources.
The radiative age of the inner source structure, estimated from
our observations, is only $\sim$ 170 yr.\\
Assuming that the radiative age and the dynamic age of the
source are similar, we predict that the parsec--scale lobes in
3C\,317 are moving away from each other at a speed 
$v_{sep}\sim 0.14 ~hc$.\\

On the arcsecond scale, the nucleus of 3C\,317 is surrounded by an 
extended amorphous ``halo'' of low brightness radio emission, which 
spatially anticorrelates with the cluster X--ray emission. 
In particular, the radio plasma is found to fill gaps of X--ray brightness, 
supporting the recent suggestion that the radio plasma sweeps the 
thermal intracluster gas and compresses it into bright shells.
An intriguing possibility is that the extended amorphous halo
is the result of reacceleration processes due to MHD turbulence in
the cooling flow.\\
The young age estimated for the radio nucleus of 3C\,317, coupled
with the existence of large scale radio emission, leads us to suggest
that the radio galaxy is characterised by intermittent radio emission
in the nucleus, and that a new phase of activity has just 
started. 
This result is relevant not only in the light of our
understanding of radio source birth and evolution, but it is also
an important piece of information for our knowlegde of the interaction
between the non thermal radio emission and the hot thermal gas in
clusters of galaxies.

\begin{acknowledgements}
The authors wish to thank M. Murgia and C. Stanghellini for their 
insightful comments on the paper.
Thanks are due to the staff of observatories who participated in 
these observations and the Socorro correlator staff. 
The National Radio Astronomy Observatory (NRAO) is operated by Associated 
Universities Inc., under cooperative agreement with the National Science 
Foundation.
This research has made use of the NASA/IPAC Extragalactic Database (NED) 
which is operated by the Jet Propulsion Laboratory, Caltech, under contract 
with NASA. This research has also made use of NASA's Astrophysics Data 
System Abstract Service.
FS acknowledges financial support from IRA-CNR under grant N. 126.59.BO.2. 
DD acknowledges financial support under grant MIUR COFIN 2002--02--8118.    
\end{acknowledgements}

\end{document}